\RequirePackage{fix-cm}
\documentclass[twocolumn]{svjour3}          
\smartqed  
\usepackage{graphicx}
\usepackage{siunitx}
\usepackage{amsmath}
\usepackage{epstopdf}
\usepackage[misc,geometry]{ifsym} 
\journalname{Granular Matter}
\begin{document}

\title{Tensile Stress Relaxation in Unsaturated Granular Materials
}


\author{Filippo Bianchi\and Marcel Thielmann\and Roman Mani\and Dani Or\and Hans J\"urgen Herrmann}


\institute{Filippo Bianchi (\Letter)\and Hans J\"urgen Herrmann \at
              Institute for Building Materials, ETH Z\"urich\\
              8093 Zurich, Switzerland\\
              \email{fbianchi@ifb.baug.ethz.ch}
           \and
           Marcel Thielmann \at
              Bayerisches Geoinstitut, University of Bayreuth\\
              95440 Bayreuth, Germany
           \and
           Dani Or \at
           	  Institute of Biogeochemistry and Pollutant Dynamics, ETH Z\"urich\\
           	  8092 Zurich, Switzerland
}

\date{Received: date / Accepted: date}

\maketitle

\begin{abstract}
The mechanics of granular media at low liquid saturation levels remain poorly understood. Macroscopic mechanical properties are affected by microscale forces and processes, such as capillary forces, inter-particle friction, liquid flows, and particle movements. An improved understanding of these microscale mechanisms is important for a range of industrial applications and natural phenomena (e.g. landslides). This study focuses on the transient evolution of the tensile stress of unsaturated granular media under extension. Experimental results suggest that the stress state of the material evolves even after cessation of sample extension. Moreover, we observe that the packing density strongly affects the efficiency of different processes that result in tensile stress relaxation. By comparing the observed relaxation time scales with published data, we conclude that tensile stress relaxation is governed by particle rearrangement and fluid redistribution. An increased packing density inhibits particle rearrangement and only leaves fluid redistribution as the major process that governs tensile stress relaxation.
\end{abstract}

\section{Introduction}
\label{intro}

Tensile strength of granular materials, namely their resistance to deformation by tensile forces, has been experimentally studied using different techniques (see \cite{Kristensen1985,Lu2007,Pierrat1998,Pierrat1997,Schubert1975,Takenaka1981,Turba1964}). The different studies show that there are several parameters that affect this property. First, the degree of liquid saturation significantly alters the tensile strength. At low saturation degrees, the liquid forms capillary bridges between the particles \cite{Iveson2001,Scheel2008NM}, that exert a cohesive force on the particles and thus increase the effective strength of a granular medium \cite{Fisher1926,Haines1925,Herminghaus2005,Hornbaker1997,Lu2007,Mitarai2006,Scheel2008JP,Schiffer2005}. At higher saturation degrees, liquid is also present in clusters, which also exert a cohesive force on the adjacent particles \cite{Iveson2001,Scheel2008NM}. At even higher saturation degrees, the abovementioned effects vanish and the strength of a granular medium decreases significantly \cite{Lu2007,Mitarai2006}.

Additionally, the effects of fluid saturation are altered by particle size and degree of compaction \cite{Iveson2001,Kim2003}. Generally, media characterised by smaller particle sizes (fine-grained materials) have a higher strength than the ones composed of larger grains (coarse-grained materials) \cite{Lu2007}. This phenomenon can be related to the increasing number of contacts between particles with decreasing particle size \cite{Carr1967}. Additionally, the effect of liquid saturation on tensile strength is different in fine- and coarse-grained materials: whereas fine-grained materials exhibit a strength peak at intermediate saturations \cite{Lu2007,Takenaka1981}, the strength of coarse-grained materials increases with increasing saturation level \cite{Lu2007,Schubert1975}. The last major property that affects tensile strength is the packing density. As one would expect, tensile strength increases with increasing packing density \cite{Kim2003,Rumpf1962}. As in the case of capillary bridges, this behaviour can be attributed to the increased number of contacts at higher packing densities \cite{Kim2003}.

Most studies mentioned above have focused on the ultimate tensile strength of an unsaturated (therefore with a liquid saturation level lower than 100~\%) granular medium, which is the highest stress one measures before the material fails. However, in a deforming granular medium, there are a number of processes that affect its stress state. First, due to stretching of the capillary bridges, their Laplace pressure changes, which in turn results in fluid exchange between bridges with differing Laplace pressure and thus also in a change in cohesive forces \cite{Herminghaus2005,Kohonen2004,Mani2015,Scheel2008JP}. Second, if a capillary bridge is sufficiently stretched, it will rupture and thus its cohesive contribution to tensile strength will be lost. Third, the particles in a granular medium may rearrange, which can alter i) their number of contacts and ii) the length of the respective capillary bridges. We expect the tensile stresses of a granular medium to vary with time due to interplay among the phenomena above, and to relieve in a way similar to viscoelastic materials (see e.g. \cite{Francois2012}). For this reason, we conducted a series of experiments where we investigated the temporal evolution of the tensile stress of an unsaturated granular medium under extension. Here we present the experimental setup and the results of the experiments, which show that the temporal stress evolution behaves differently for different particle packings. We then discuss the possible causes for the behaviours observed in the experiments.

\section{Experiments}
\label{sec:Experiments}

\subsection{Experimental setup}
\label{sec:Experimental setup}

\begin{figure}
\includegraphics[width=\columnwidth]{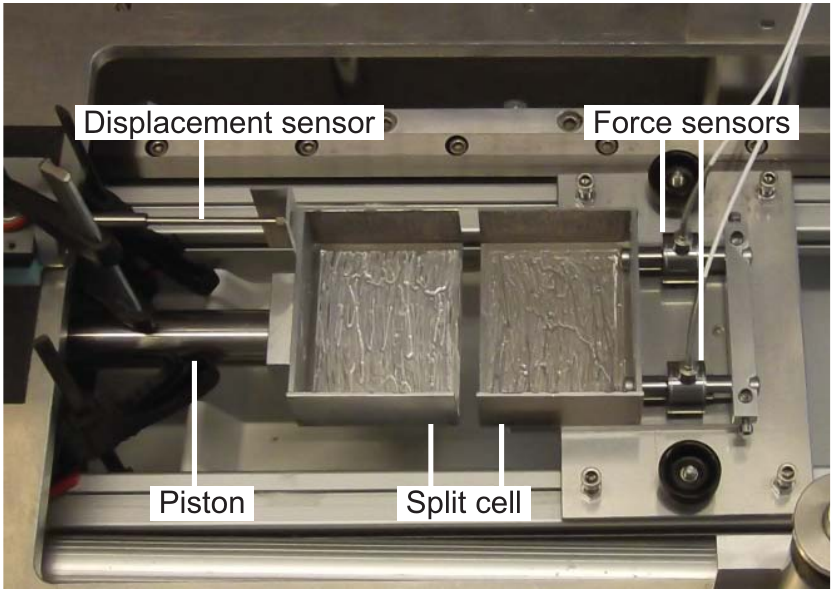}
\caption{Experimental setup}
\label{Setup}
\end{figure}

The experimental setup consists of a split cell with dimensions 9.6$\times$15$\times$3.6~\si{\centi\metre^3} (see Fig. \ref{Setup}). Cell surfaces are roughened to increase friction and prevent the grains from sliding against them. The left half of the cell is connected to a piston, which can be moved at arbitrary speeds. Its movement is measured with a displacement sensor (Inelta ISDT50). The right part of the cell is connected to two force sensors (Interface WMCFP, 500~\si{\gram} capacity) and is mounted on bearings. Displacement sensor and force sensors are connected to a datalogger (Campbell Scientific CR1000), while the piston is controlled by a displacement controller (PERO GmbH). The overall measurement accuracy is $\pm$18~\si{\micro\metre} for displacement measurements and $\pm$0.02~\si{N} for force measurements. A similar setup has already been used by Ref. \cite{Pierrat1997}.

\begin{figure}
\includegraphics[width=\columnwidth]{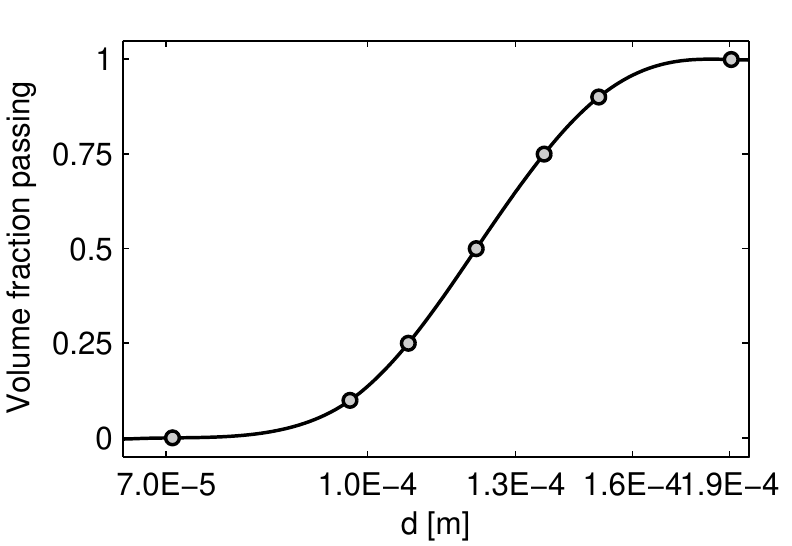}
\caption{Cumulative grain size distribution of the glass beads; $d$ is the grain diameter}
\label{GrainSizeSI}
\end{figure}

The granular material consists of glass beads with mean grain radius of 61 \si{\micro\metre} (see Fig. \ref{GrainSizeSI} for grain size distribution).
The glass beads are wetted with either water or a mixture of water and Triton X-100 (Sigma Aldrich) with a molarity of 1.5$\cdot$10\textsuperscript{-4}~\si{M}. Triton X-100 was added to manipulate the surface tension, this surfactant reduces the surface tension from 72.8$\cdot$10\textsuperscript{-3}~\si{N/m} (pure water) to 32.8$\cdot$10\textsuperscript{-3}~\si{N/m} \cite{Labajos2006}. Both surface tension values refer to the fluid as it is prepared before mixing it with the glass beads.

\subsection{Sample preparation}
\label{sec:Sample preparation}

To investigate the influence of packing density, two types of samples were used: the first type had a low packing density (with a solid fraction $\Phi$~$\simeq$~0.41), the second had a slightly higher packing density ($\Phi$~$\simeq$~0.51). Packing density was measured by weighing the granular material before filling the split cell and by estimating its volume after placing it inside the split cell. Using a glass bead density of 2.5$\cdot$10\textsuperscript{3}~\si{kg/m^3}, we then estimated the packing density by computing the total glass bead volume and dividing it by the total cell volume.

The wet granular mixture was prepared by adding the desired amount of liquid to the glass beads. The two phases were mixed and left to equilibrate for at least 30 minutes to ensure a homogeneous fluid distribution. In all experiments presented here, the volumetric liquid content ($\theta_v$, defined as liquid volume divided by total sample volume) is lower than  2.4~\%, thus the sample was in the pendular regime. In this regime, liquid is retained solely in capillary bridges \cite{Scheel2008NM,Scheel2008JP}.

Low density samples were prepared by filling the split cell with a funnel without further compaction. The sample surface was levelled by removing the exceeding material. The packing fraction of those samples was 0.41~$\pm$0.02. Samples with higher packing fraction were prepared following the same procedure, but were additionally vibrated with a frequency of 150~\si{\hertz}. The material was vibrated from the top via a vibration conductor made of extruded rigid polystyrene (SAGEX XPS Styro by Sager AG). The vibrating device on top consisted of a loud speaker connected to an oscilloscope. Since it was not possible to precisely set a vibration amplitude (which was most probably modified by the vibration conductor as well), we are not able to give an accurate information on this parameter. Samples were vibrated for 30 seconds. The packing fraction of those samples was 0.51~$\pm$0.02. To reduce evaporation the cell was covered after sample preparation.

\subsection{Experimental procedure}
\label{sec:Experimental procedure}

To investigate the temporal tensile stress evolution of the wet granular sample, we extended the samples at a velocity of 4$\cdot$10\textsuperscript{-6}~\si{m/s} until reaching the peak strength of the sample. The strain $\epsilon=\Delta l/l$ (where $\Delta l$ is the displacement of the left cell and $l$ is the length of the sample before extension) at which peak strength is reached is termed the \textit{critical strain} \cite{Mitarai2006}. In our experiments, the critical strain was reached at displacements of 1.5 to 2.3 bead radii. Being the cell covered during the experiments, we did not evaluate local strains inside the samples. Extension was then halted for 30 minutes. After this phase extension was resumed until no more tensile strength was exerted by the material. During the entire experiment, the force ($F$) and displacement ($\Delta l$) were recorded at 1~\si{\hertz} sampling rate.

\section{Results}
\label{sec:Results}

\begin{figure}[h!]
\includegraphics[width=\columnwidth]{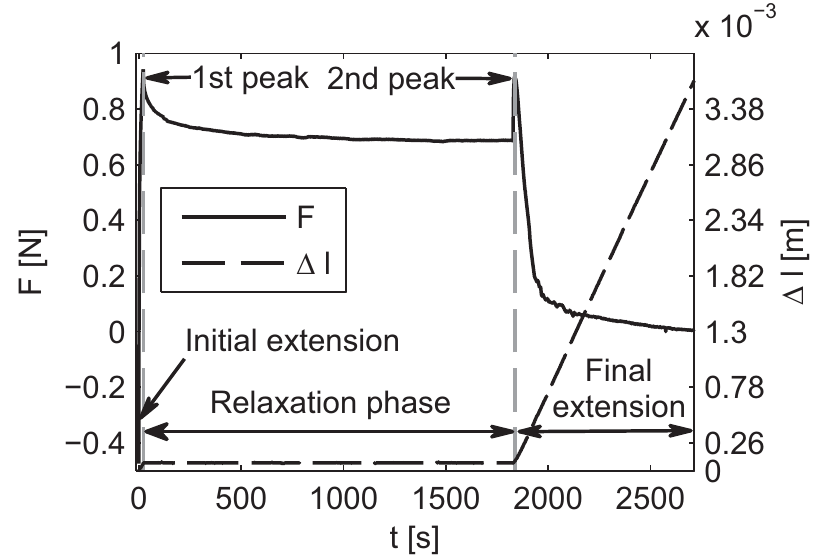}
\caption{Temporal evolution of force and displacement in a typical experiment ($\Phi$~$\simeq$~0.51, $\Gamma$~=~32.8$\cdot$10\textsuperscript{-3}~\si{N/m}); force data is corrected for intrinsic friction in the setup; $t$ is the elapsed time}
\label{Example}
\end{figure}

In Fig. \ref{Example} we show the evolution of force and displacement for a typical experiment (experiment 7 of Table \ref{Compact}).

The experiment can be divided into three phases: initial extension, relaxation, and final extension. The different stages are characterised by the evolution of the measured force. During the initial extension phase, the force increases until reaching peak strength, which corresponds to the material ultimate tensile strength. Peak strength is defined by looking at force data recorded by sensors. In loose packing experiments run with pure water as interstitial fluid, the initial extension is manually stopped (acting on the piston displacement controller) as soon as the sensors show that the force is decreasing, thus right after the tensile strength of the material has been reached. The peak force value occurs at displacements of 9$\cdot$10\textsuperscript{-5} to 1.4$\cdot$10\textsuperscript{-4}~\si{\metre} and corresponds to the highest force that can be sustained by the sample. In the other cases, extension is stopped by halting the piston movement at the fixed displacement of 1.5 mean bead radii (9$\cdot$10\textsuperscript{-5}~\si{\metre}). Thus the force peak corresponds to the force recorded by the sensors when the left cell has been displaced by 9$\cdot$10\textsuperscript{-5}~\si{\metre}. In the following phase the force relaxes (this phase is consequently named the \textit{relaxation phase}). After 30 minutes extension is resumed. During this phase, we observe an increase in force until a second peak is reached. Note that this force peak is lower than after the first extension phase. Subsequently, the measured force decreases rapidly, because a fracture fully develops inside the sample, so that the grains on the left cell are completely disconnected from the ones on the right cell. At the end of the final extension phase the two sides of the sample are completely disconnected. The remaining force can be related to the friction between the sample and the bearings.

Experiments were run with loose packing samples (experiments 1 to 4 of Table \ref{Loose}) and vibrated samples (experiments 5 to 7 of Table \ref{Compact}). The volumetric liquid content was 1.5~\% in loose packing samples and 1.8~\% in vibrated samples. Interstitial liquid of lower packing density experiments was pure water (surface tension: $\Gamma$~=~72.8$\cdot$10\textsuperscript{-3}~\si{N/m}), while water with the addition of Triton X-100 was used in the case of vibrated samples ($\Gamma$~=~32.8$\cdot$10\textsuperscript{-3}~\si{N/m}). Two different fluids were used to observe if surface tension had a relevant effect on stress relaxation. To discern the effect of surface tension from the one of packing density, pure water was used in one vibrated packing experiment and water with the addition of Triton X-100 was used in one loose packing experiment.

\begin{figure}[h!]
\includegraphics[width=\columnwidth]{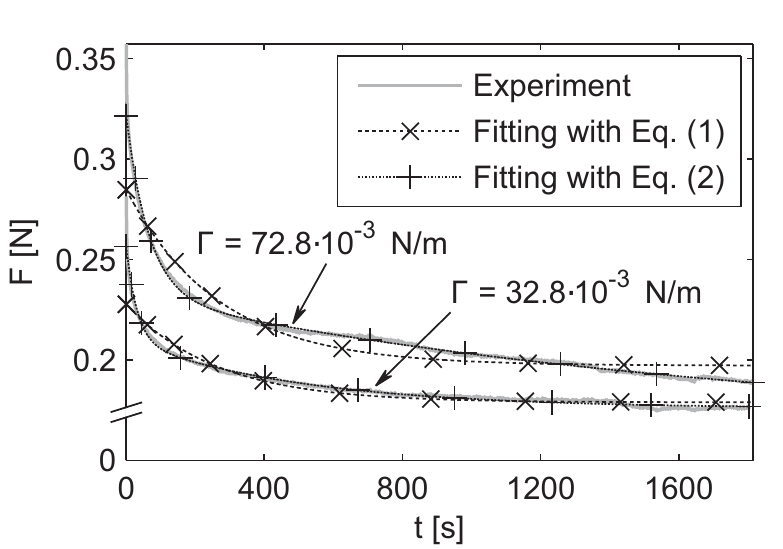}
\caption{Force decay after stopping the initial extension for two loose samples; experimental data and curve fitting; $\Phi$~$\simeq$~0.41; force data is corrected for intrinsic friction in the setup}
\label{Loose_exp_2}
\end{figure}

\begin{figure}[h!]
\includegraphics[width=\columnwidth]{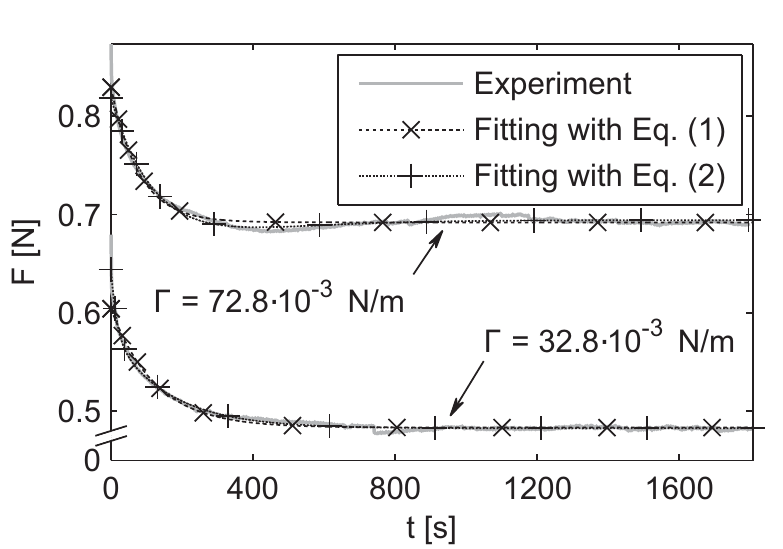}
\caption{Force decay after stopping the initial extension for two vibrated samples; experimental data and curve fitting; $\Phi$~$\simeq$~0.51; force data is corrected for intrinsic friction in the setup}
\label{Vib_exp_5}
\end{figure}

Our analysis concentrates on the relaxation phase (see Fig. \ref{Example}). Initially, relaxation is very fast and then slows down. At the end of the relaxation phase we observe a steady state (see Fig. \ref{Loose_exp_2} and \ref{Vib_exp_5}).

Although the scatter in peak force of the different experiments is large (standard deviations of 0.022~\si{N} and 0.119~\si{N} for loose packing and vibrated packing experiments respectively, with mean peak force values of 0.332~\si{N} and 0.845~\si{N}, see Tables \ref{Loose} and \ref{Compact}), we observe two distinct types of force relaxation (see also Fig. \ref{Loose_exp_2} and Fig. \ref{Vib_exp_5}). To better characterise the two relaxation types, the force relaxation curve is fitted using two different functions: a single exponential decay function \eqref{oneexp} and a double exponential decay function \eqref{twoexp}.
\begin{eqnarray}\label{oneexp}
\displaystyle F&=&a_1e^{-t/\tau_1}+b_1\\\label{twoexp}
\displaystyle F&=&a_2e^{-t/\tau_2}+a_3e^{-t/\tau_3}+b_2
\end{eqnarray}
$t$ is the time elapsed since the initial extension is stopped; $\tau_1$, $\tau_2$, and $\tau_3$ are the relaxation times of the decay; $a_1$, $b_1$, $a_2$, $a_3$, and $b_2$ are fitting parameters.

In Fig. \ref{Loose_exp_2}, we show the force relaxation curve of experiment 2 (with a packing fraction of $\simeq$~0.41 and a surface tension of 72.8$\cdot$10\textsuperscript{-3}~\si{N/m}) together with the fitting curves of both decay functions. The lower curves show an experiment run with the same packing fraction, but with a fluid with lower surface tension (32.8$\cdot$10\textsuperscript{-3}~\si{N/m}). We observe that the double exponential decay function fits the relaxation curve significantly better (a triple exponential decay does not significantly improve the fit). This improvement can also be appreciated by looking at the goodness of fit of decay functions (see Table \ref{Gof}). The SSE (sum of squares due to error) is reduced on average by a factor of ten when fitting with Eq. \eqref{twoexp} instead of Eq. \eqref{oneexp}.

Fig. \ref{Vib_exp_5} shows the force relaxation curve of a vibrated experiment (experiment 5, $\Phi$~$\simeq$~0.51, $\Gamma$~=~32.8$\cdot$10\textsuperscript{-3}~\si{N/m}) and the two fitting curves. The higher curves show a vibrated experiment run with an interstitial fluid with higher surface tension (72.8$\cdot$10\textsuperscript{-3}~\si{N/m}). We observe that both decay functions fit the experimental data equally well and that no substantial improvement is obtained using a double exponential decay fitting function. When looking at SSE of decay functions (Table \ref{Gof}), we also notice that the improvement obtained with Eq. \eqref{twoexp} is not so relevant as in loose samples (SSE is reduced on average by a factor of four instead of ten).

The values of relaxation times computed for the different experiments are reported in Table \ref{Loose} (loose packing experiments) and Table \ref{Compact} (vibrated experiments), while Table \ref{Gof} shows the goodness of fit.

\begin{table}[h!]
\caption{Loose packing experiments: force values at the first peak (F\textsubscript{1st peak}) and at the end of the relaxation phase (F\textsubscript{end}) and relaxation times; $\Phi$~$\simeq$~0.41, $\theta_v$~=~1.5~\%, $\Gamma$ of fluid~=~72.8$\cdot$10\textsuperscript{-3}~\si{N/m}; $\overline{x}$ are the arithmetic means; $\sigma$ are the standard deviations; $\frac{x_{max}}{x_{min}}$ are the ratios between the highest and the lowest values}
\label{Loose}
\begin{tabular}{llllll}
\hline\noalign{\smallskip}
Exp. &  F\textsubscript{1st peak} & F\textsubscript{end} & $\tau_1$ [s] & $\tau_2$ [s] & $\tau_3$ [s]\\
& [N] & [N] & & &\\
\noalign{\smallskip}\hline\noalign{\smallskip}
1 & 0.32 & 0.16 & 457 & 54 & 735\\
2 & 0.36 & 0.19 & 268 & 66 & 1825\\
3 & 0.30 & 0.15 & 354 & 30 & 612\\
4 & 0.35 & 0.23 & 284 & 61 & 1665\\
\noalign{\smallskip}\hline\noalign{\smallskip}
$\overline{x}$ & 0.33 & 0.18 & 340.6 & 52.6 & 1209.0\\
$\sigma$ & 0.03 & 0.04 & 74.4 & 13.8 & 540.6\\
$\frac{x_{max}}{x_{min}}$ & 1.20 & 1.53 & 1.7 & 2.2 & 3.0\\
\noalign{\smallskip}\hline
\end{tabular} 
\end{table}

\begin{table}[h!]
\caption{Vibrated experiments: force values at the first peak and at the end of the relaxation phase and relaxation times; $\Phi$~$\simeq$~0.51, $\theta_v$~=~1.8~\%, $\Gamma$ of fluid~=~32.8$\cdot$10\textsuperscript{-3}~\si{N/m}}
\label{Compact}
\begin{tabular}{llllll}
\hline\noalign{\smallskip}
Exp. &  F\textsubscript{1st peak} & F\textsubscript{end} & $\tau_1$ [s] & $\tau_2$ [s] & $\tau_3$ [s]\\
& [N] & [N] & & &\\
\noalign{\smallskip}\hline\noalign{\smallskip}
5 & 0.68 & 0.48 & 125 & 16 & 163\\
6 & 0.90 & 0.68 & 104 & 10 & 146\\
7 & 0.96 & 0.70 & 207 & 41 & 400\\
\noalign{\smallskip}\hline\noalign{\smallskip}
$\overline{x}$ & 0.85 & 0.62 & 145.0 & 22.2 & 236.5\\
$\sigma$ & 0.15 & 0.12 & 44.7 & 13.5 & 115.9\\
$\frac{x_{max}}{x_{min}}$ & 1.41 & 1.46 & 2.0 & 4.3 & 2.7\\
\noalign{\smallskip}\hline
\end{tabular} 
\end{table}

\begin{table}[h!]
\caption{Goodness of fit of loose packing and vibrated experiments; SSE \eqref{oneexp} is the sum of squares due to error, computed using Eq. \eqref{oneexp} as fitting function; SSE \eqref{twoexp} is the sum of squares due to error, computed using Eq. \eqref{twoexp} as fitting function; $\overline{SSE}$ are the arithmetic means}
\label{Gof}
\begin{tabular}{llllll}
\hline\noalign{\smallskip}
\multicolumn{3}{l}{Loose packings} & \multicolumn{3}{l}{Vibrated packings}\\
\hline\noalign{\smallskip}
Exp. &  SSE \eqref{oneexp} & SSE \eqref{twoexp} & Exp. & SSE \eqref{oneexp} & SSE \eqref{twoexp}\\
& [N\textsuperscript{2}] & [N\textsuperscript{2}] & & [N\textsuperscript{2}] & [N\textsuperscript{2}]\\
\noalign{\smallskip}\hline\noalign{\smallskip}
1 & 3.1$\cdot$10\textsuperscript{-2} & 3.1$\cdot$10\textsuperscript{-3} & 5 & 2.3$\cdot$10\textsuperscript{-2} & 0.8$\cdot$10\textsuperscript{-2}\\
2 & 7.3$\cdot$10\textsuperscript{-2} & 3.9$\cdot$10\textsuperscript{-3} & 6 & 5.6$\cdot$10\textsuperscript{-2} & 2.1$\cdot$10\textsuperscript{-2}\\
3 & 3.7$\cdot$10\textsuperscript{-2} & 5.6$\cdot$10\textsuperscript{-3} & 7 & 8.6$\cdot$10\textsuperscript{-2} & 1.0$\cdot$10\textsuperscript{-2}\\
4 & 3.5$\cdot$10\textsuperscript{-2} & 4.6$\cdot$10\textsuperscript{-3} &  &  & \\
\noalign{\smallskip}\hline\noalign{\smallskip}
$\overline{SSE}$ & 4.4$\cdot$10\textsuperscript{-2} & 4.3$\cdot$10\textsuperscript{-3} &  & 5.5$\cdot$10\textsuperscript{-2} & 1.3$\cdot$10\textsuperscript{-2}\\
\noalign{\smallskip}\hline
\end{tabular} 
\end{table}

\begin{figure}[h!]
\includegraphics[width=\columnwidth]{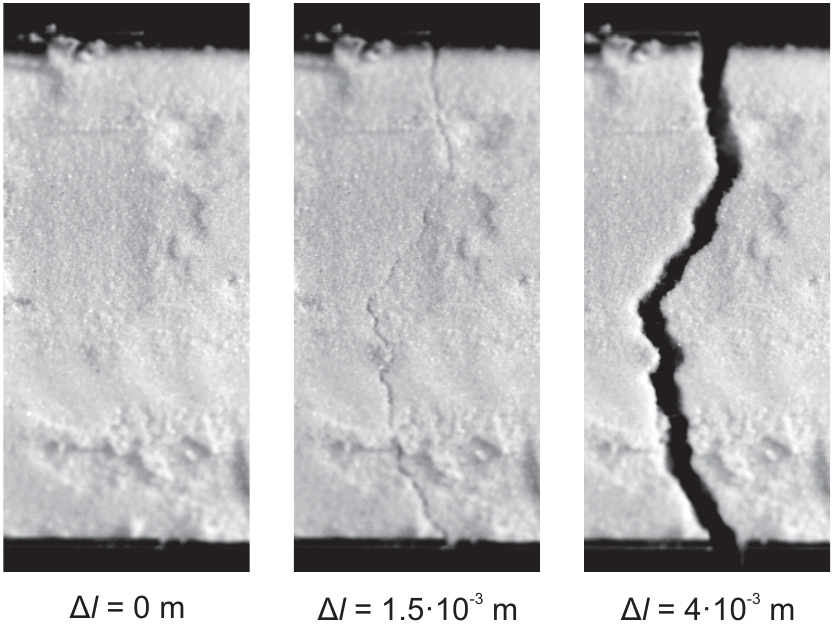}
\caption{Fracture evolution in a loose packing experiment; the fracture becomes visible when $\Delta l$~=~1.5$\cdot$10\textsuperscript{-3}~\si{m}}
\label{Crack_loose}
\end{figure}

\begin{figure}[h!]
\includegraphics[width=\columnwidth]{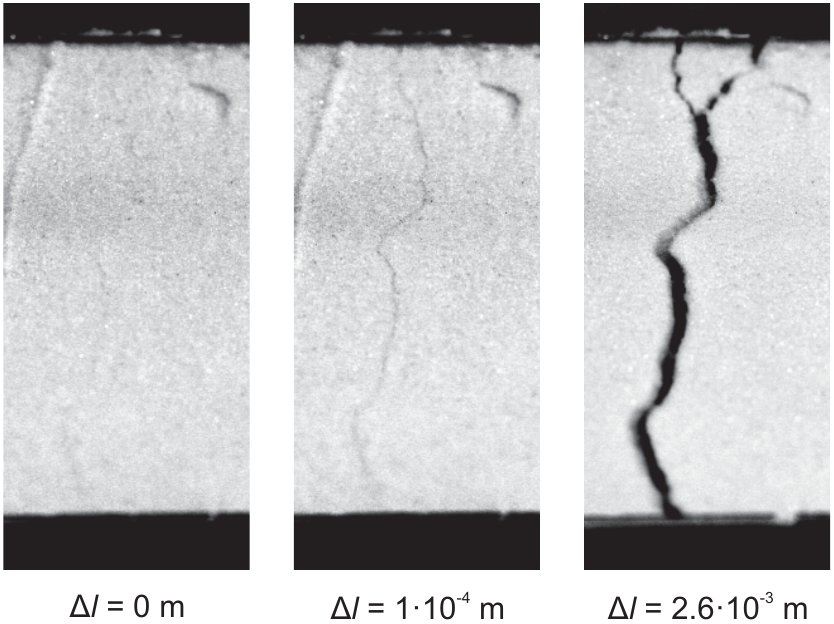}
\caption{Fracture evolution in a vibrated packing experiment; the fracture becomes visible when $\Delta l$~=~1$\cdot$10\textsuperscript{-4}~\si{m}}
\label{Crack_vibrated}
\end{figure}

To better observe the formation of a rupture surface, we also ran experiments where the sample was not covered. As evaporation was significantly larger in those cases, we cannot directly compare the force curves to the experiments presented above. However, depending on the packing of the sample, we could qualitatively observe that the formation of a rupture surface occurred much later in loosely packed samples than in close packed samples. As shown in Fig. \ref{Crack_loose}, when using loose packing samples ($\Phi$~$\simeq$~0.41), the fracture developed during the final extension phase. The material had to reach a strain as high as $\epsilon$~$\simeq$~0.01 before a rupture surface was formed. In the vibrated case ($\Phi$~$\simeq$~0.51), the rupture surface appeared much earlier (see Fig. \ref{Crack_vibrated}). One could clearly observe its formation already at a strain of 6.7$\cdot$10\textsuperscript{-4} (end of initial extension phase).

\section{Discussion}
\label{sec:Discussion}

Our results illustrate that the tensile stress of an unsaturated granular material is time dependent. We observe stress relaxation, which can be fitted with either a single or a double exponential decay.

Loose packing samples exhibit a force decay which can be well fitted using two relaxation times. This becomes evident when looking at Table \ref{Gof}. When a single exponential decay function is used to fit the experimental data, SSE values vary between 3.1$\cdot$10\textsuperscript{-2}~\si{N^2} and 7.3$\cdot$10\textsuperscript{-2}~\si{N^2}. These errors are reduced by one order of magnitude if a double exponential function is used (3.1$\cdot$10\textsuperscript{-3}~\si{N^2} to 5.6$\cdot$10\textsuperscript{-3}~\si{N^2}). Eq. \eqref{twoexp} improves the fitting of vibrated experiments as well, but in a less relevant way compared to loose packings. The errors obtained fitting with Eq. \eqref{oneexp}, range from 2.3$\cdot$10\textsuperscript{-2}~\si{N^2} to 8.6$\cdot$10\textsuperscript{-2}~\si{N^2}, while the ones obtained with Eq. \eqref{twoexp} range between 0.8$\cdot$10\textsuperscript{-2}~\si{N^2} and 2.1$\cdot$10\textsuperscript{-2}~\si{N^2}. Due to sample variability (even within the same experimental conditions), the improvement obtained using Eq. \eqref{twoexp} can also be relevant for vibrated experiments (the ratio between SSE obtained with a single and a double exponential decay is as high as 8.5 for experiment 7). Yet it should be remarked, that an increase in the number of fitting parameters results most likely in any case in a reduction of SSE. When looking at the whole set of data (Table \ref{Gof}), a clear difference emerges between loose and vibrated experiments, in respect that Eq. \eqref{twoexp} is needed to fit loose packing experiments, while Eq. \eqref{oneexp} is able to fit vibrated experiments.

It becomes apparent that the force decay type (force variation as a function of time) depends on the packing of the sample, while liquid surface tension only affects the strength values. Sample packing then determines the processes governing force relaxation. The number of relaxation times (one or two) suggests that either one or two phenomena are dominant.

\begin{figure*}
\includegraphics[width=\textwidth]{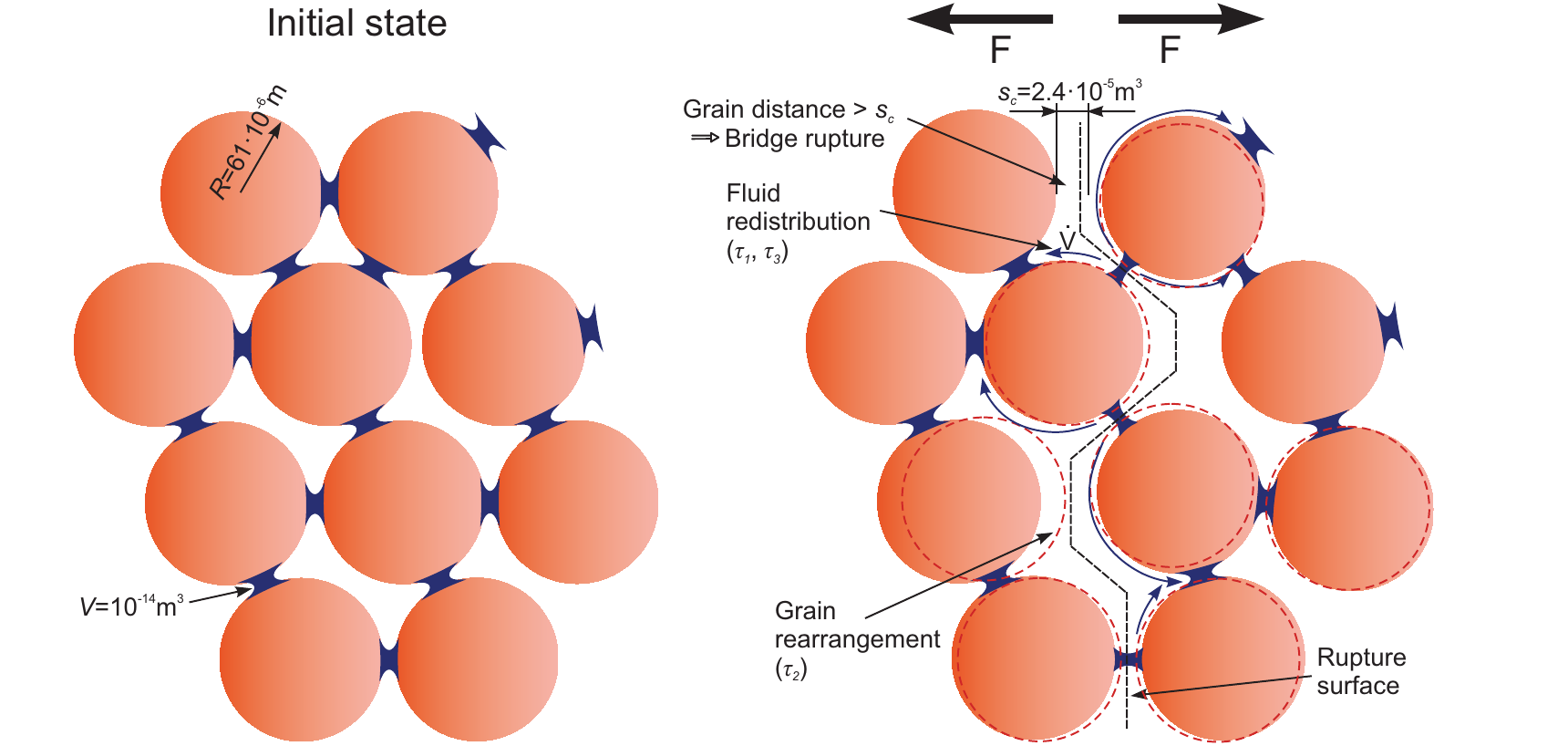}
\caption{Bidimensional sketch of the different processes occurring in the granular medium during extension; out of plane bridges and grains are not depicted for clarity}
\label{draw}
\end{figure*}

Three main phenomena are expected to occur during the relaxation phase (see Fig. \ref{draw}): i) liquid bridge ruptures \cite{Kohonen2004,Willett2000}, ii) fluid redistribution \cite{Herminghaus2005,Kohonen2004,Mani2015,Scheel2009,Scheel2008JP}, and iii) grain rearrangement \cite{Hartley2003,Utter2004}.

We estimate the critical bridge length, i.e. the distance between two beads at which the bridge pinches off, to evaluate the importance of liquid bridge ruptures during the relaxation phase. An empirical expression used to calculate the critical bridge length has been proposed by Ref. \cite{Willett2000} and is given by
\begin{equation}\label{sc}
\displaystyle s_c=R\left(1+\frac{\theta}{2}\right)\left(\frac{\sqrt[3]{V}}{R}+\frac{\sqrt[3]{V^2}}{10R^2}\right)
\end{equation}
where $s_c$ is the critical bridge length, $V$ is the bridge volume, $R$ is the mean bead radius (61$\cdot$10\textsuperscript{-6}~\si{\metre}), and $\theta$ is the contact angle of the fluid at the grain surface (which is assumed to be 0~\si{\degree} in this calculation). To estimate $s_c$ we calculate the mean bridge volume of the sample. This is done taking into account the weight of the beads composing the sample (0.5~\si{kg} for loose packing experiments, 0.6~\si{kg} for vibrated experiments), their density (2.5$\cdot$10\textsuperscript{3}~\si{kg/m^3}), and their diameter. One can thus estimate the total number of beads inside the split cell and the total number of liquid bridges (we assume that, on average, six bridges are connected to every particle \cite{Kohonen2004}). Knowing the amount of water added to the medium (7.5~\si{ml} in loose packing experiments, 9~\si{ml} in vibrated experiments), the bridge volume can be estimated to be about 10\textsuperscript{-14}~\si{m^3} in both loose packing experiments and vibrated experiments. The critical length for a bridge with this volume is 2.4$\cdot$10\textsuperscript{-5}~\si{\metre}. This should be reached within 6~\si{s} during the initial extension phase (at the force peak, the left cell displacement is at least 9$\cdot$10\textsuperscript{-5}~\si{\metre}). If the particles were firmly connected to the cell and if there were no particle movement allowed, liquid bridges across the gap section would stretch as much as the sample. Hence they would break during the initial extension phase, and bridge ruptures would not play any significant role during the relaxation phase. However granular packings are not perfectly rigid, but they can deform. Therefore some bridges across the gap section are expected to survive initial extension. These bridges will contribute to material tensile strength also during the relaxation phase.

Fluid redistribution among capillary bridges occurs due to differences in their Laplace pressure. The liquid moves through thin wetting layers surrounding the grains and through the vapour phase \cite{Herminghaus2005,Kohonen1999,Seeman2001}. During extension, the bridges in the area of contact between the two sides of the cell are elongated, and thus their Laplace pressure increases \cite{Lambert2008,Lian2016,Lian1993}. This increase results in a fluid flux away from the extended region of the sample. This flux decreases the volume of the elongated bridges, until either they equilibrate their pressure with the other bridges or they rupture. The result of this fluid depletion is a temporal reduction in material internal stress. The force exerted by liquid bridges decreases when their volume reduces, as can be seen in the empirical expression proposed by Ref. \cite{Willett2000}
\begin{equation}\label{willett}
\displaystyle F_{bridge}=\frac{2\pi R\Gamma\cos\theta}{1+1.05s\sqrt{\dfrac{R}{V}}+2.5s^2\dfrac{R}{V}}
\end{equation}
where $F_{bridge}$ is the bridge strength, and $s$ is the bridge length. The process of pressure equilibration through fluid redistribution is rather slow. According to Ref. \cite{Herminghaus2005}, the relaxation time of liquid redistribution after the end of any perturbation of the medium is about 300~\si{s} (for water at room temperature), and it scales with $\mu R^2$ ($\mu$ is the dynamic viscosity of the liquid). The author of Ref. \cite{Scheel2009} has also proposed that liquid redistribution relaxation time is directly proportional to liquid viscosity, but he claims it is also inversely proportional to liquid surface tension. Therefore the relaxation time of fluid redistribution should scale with $\mu/\Gamma$. Being of the order of hundreds of seconds, we expect fluid redistribution to play an important role for several minutes during the relaxation phase of our experiments.

The relative movement of particles is possible in ductile samples. With increasing compaction, such a rearrangement becomes progressively more difficult, since each particle movement involves mobilising a larger number of grains \cite{Hartley2003,Utter2004}. Grain rearrangement does not only occur when the medium is subjected to external stresses but still persists after the removal of any applied load. According to Ref. \cite{Hartley2003} the time scale of this relative movement varies between 20 and 300 s after unloading the material. The authors of Ref. \cite{Utter2004} have also suggested that this time scale should last a few tens of seconds. Thus grain rearrangement is likely to occur not only during the initial extension phase, but also during the beginning of the relaxation phase.

The different phenomena here described do not occur independently, but they are interrelated. Bridge ruptures will not only occur due to the initial extension of the medium, because packings are not completely rigid, and not all the bridges have the same volume. Thus several bridges in the area of contact between the two sides of the cell will survive the extension phase. Because of fluid redistribution between these bridges and the neighbouring one, and because of grain rearrangement, some ruptures are likely to occur also during the relaxation phase. However these ruptures (thus also the time scale) are entirely dependent on the other two phenomena. Fluid redistribution affects the strength of liquid bridges as well, thus the force and contact network inside the medium. This can lead to grain rearrangement which also modifies the internal stress state of the packing. Particle rearrangement occurring at the beginning of the relaxation phase modifies the capillary bridges network. Bridges are elongated or shortened, therefore their pressure varies and liquid flux is initiated. However, the time scales of these phenomena do not change because grain rearrangement induced by fluid redistribution stops when fluid redistribution ends, and fluid redistribution induced by grain rearrangement stops when grain rearrangement ends.

In the loose packing experiments, force relaxation can be described by a double exponential decay (Fig. \ref{Loose_exp_2}), whose characteristic relaxation times are approximately 50~\si{s} ($\tau_2$) and 1200~\si{s} ($\tau_3$) (see Table \ref{Loose}). This indicates that two different phenomena dominate during the relaxation phase. Both grain rearrangement and fluid redistribution are therefore likely to play a significant role in altering the stress state of the granular medium. Furthermore the relaxation times we obtained from experiments are comparable to grain rearrangement and fluid redistribution time scales proposed in the literature \cite{Hartley2003,Herminghaus2005,Utter2004}. Using the previous arguments, we suggest that the short relaxation time scale in our experiments can be attributed to grain rearrangement and the long relaxation time scale to fluid redistribution.

The force relaxation of vibrated packing experiments follows a single exponential decay (Fig. \ref{Vib_exp_5}). This means that only one phenomenon is dominant during the relaxation phase. Being the relaxation time of force decay ($\simeq$~150~\si{s}, see Table \ref{Compact}) comparable to the one of fluid redistribution proposed by Ref. \cite{Herminghaus2005}, we attribute the observed time scale to fluid redistribution. Grain rearrangement could also occur during the relaxation phase, but we expect that it does not play a significant role.

The observation of fracture evolution (Fig. \ref{Crack_loose}) also suggests that grain rearrangement plays an important role in loose packing experiments. The rupture surface is only visible during the second extension phase, when the displacement of the left cell is above 1~\si{mm}. This shows the high ductility of loose packing samples, in which grains are able to move with respect to each other. Vice versa, in vibrated experiments the rupture surface appears during the first extension phase (Fig. \ref{Crack_vibrated}), when the displacement of the left cell is only 0.1~\si{mm}. This indicates that the mobility of particles inside the medium is limited, and that vibrated samples are more rigid and brittle than loose packings. Grain rearrangement is therefore not expected to be an important phenomenon during the relaxation phase of vibrated experiments.

In general, the results from vibrated experiments exhibit shorter relaxation times compared to loose packing experiments (see Table \ref{Loose} and Table \ref{Compact}). This can be explained by higher bridge connectivity in close packings. As shown by Ref. \cite{Kohonen2004}, the average number of liquid bridges per grain increases with increasing packing density. Fluid depletion from the central part of the sample becomes faster as the connectivity increases. The reasons are the reduced distances among bridges (lower energy loss through wetting films) and the increased outgoing flux from high pressure bridges due to their connection to a higher number of low pressure bridges. Other parameters could also affect relaxation times, e.g. bridge volume size distributions, coordination number variability, or saturation degree. For instance, the equilibration time should increase with decreasing liquid bridge size, because of an increment of the distance among bridges. Yet local gradients in capillary pressure would also increase, leading to higher flow rates. However our experiments do not provide any accurate data for the evaluation of these parameters, hence we do not consider them in the present discussion.

Vibrated samples show a higher tensile strength than loose samples, although their interstitial liquid surface tension is lower (see Table \ref{Loose} and Table \ref{Compact}). One reason is the higher coordination number of higher density packings, thus the higher number of water bridges connected to the grains \cite{German2014,Kim2003,Kohonen2004}, which enhances sample strength. The other reason is the shorter bridge length of vibrated packings, as grains are closer to each other. As can be seen in Eq. \eqref{willett}, bridge forces increase with decreasing bridge length.

Applying Rumpf's model \cite{Rumpf1962}, we can compare the tensile strengths of our experiments with theoretical predictions. According to Rumpf's model, tensile strength of granular materials in the pendular regime could be estimated with the following expression:
\begin{equation}\label{rumpf}
\displaystyle \sigma_t=\frac{\Phi kF_{bridge}}{\pi {\left(2R\right)}^2}
\end{equation}
where $\sigma_t$ is the cohesive stress, and $k$ is the coordination number (considered to be 6 in both loose and vibrated packings \cite{Kohonen2004}). We compute $F_{bridge}$ assuming a mean bridge volume of 10\textsuperscript{-14}~\si{m^3} (as in the calculation of $s_c$). The mean bridge length $s$ is computed considering the beads uniformly displaced as in a simple cubic packing (therefore they are surrounded by 6 neighbours). Knowing the geometrical sample dimensions and the number of beads inside the packing, we can estimate the distance between the centers of two particles (132.5~\si{\micro\metre} in loose packing experiments and 123.5~\si{\micro\metre} in vibrated experiments). Subtracting the grain diameter (122~\si{\micro\metre}) from these distances, one finds the mean bridge length inside the samples to be 10.5~\si{\micro\metre} in loose packings and 1.5~\si{\micro\metre} in vibrated packings. It is thus possible to estimate $F_{bridge}$ (we assume a 0~\si{\degree} contact angle) and consequently $\sigma_t$ (457~\si{N/m^2} loose samples, 722~\si{N/m^2} vibrated samples). Tensile strength is then computed by multiplying $\sigma_t$ with the cross section of our samples (3.3$\cdot$10\textsuperscript{-3}~\si{m^2} in loose packing experiments, 3.2$\cdot$10\textsuperscript{-3}~\si{m^2} in vibrated experiments), so that we obtain a theoretical tensile strength of 1.49~\si{N} for loose samples and 2.29~\si{N} for vibrated samples. According to Pierrat and Caram \cite{Pierrat1997} Eq. \eqref{rumpf} overestimates tensile strength by a factor of approximately 1.5, because it considers isostatic stress conditions in the granule instead of uniaxial stress conditions. The theoretical tensile strength for our samples could thus be more precisely estimated as 0.99~\si{N} for loose packings and 1.53~\si{N} for vibrated packings. These values are about 2 to 3 times larger than our experimental observations (see Table \ref{Loose} and Table \ref{Compact}). However it should be mentioned that the estimation of $\sigma_t$ strongly depends on the bridge length, therefore any inaccuracy in the calculation of such a parameter can lead to very different results. We therefore believe that the discrepancy between our experiments and theoretical strength is mostly due to inaccuracies in the estimation of calculation parameters.

A heterogeneous stress state of the granular sample in the separation region of the split cell could also explain the discrepancy between theoretical strength and our experiments. A fracture develops in the middle of the sample during the experiments, as can be seen in Fig. \ref{Crack_loose} and Fig. \ref{Crack_vibrated}. The fracture could appear shortly after initial extension is started, and it would be invisible from above until it reaches the top of the packing. Similar to brittle materials, stresses could then concentrate at the tip of the fracture and thus reduce sample strength. The work needed to create the fracture surface can be calculated by integrating the force over the displacement of the initial extension phase. Depending on the experiment, this work ranges between 2.4$\cdot$10\textsuperscript{-5} and 5.0$\cdot$10\textsuperscript{-5}~\si{J}, and it comprises both the work required to break liquid bridges and the dissipated work due to particle movements. This irreversible work depends on liquid redistribution among capillary bridges (especially among those close to the fracture tip). The tensile strength of an unsaturated granular medium could thus be related to the work needed to break liquid bridges, providing a new approach for tensile strength estimation. This approach would be complementary to Rumpf's model, which is based on the calculation of the bonding force of uniformly distributed capillary bridges.

Most of the previously described processes occur also in sheared granular materials. In this case, liquid bridges inside shear bands are elongated due to the sliding of grains. If $s_c$ is reached, bridges will pinch off and their fluid will be redistributed \cite{Mani2013,Mani2012}. If $s_c$ is not reached, fluid depletion from shear bands will occur due to the pressure rise of elongated bridges. Therefore we expect stress relaxation to occur also in sheared unsaturated granular materials. In nature both shear and tension coexist at the same time. For instance in landslides, shear occurs at the interface between the landslide body and the bedrock. Traction occurs at the crown.

\section{Conclusions}
\label{sec:Conclusions}

We have studied the tensile stress of partially saturated granular media in the pendular regime. We have shown that tensile stress decreases once extension is stopped. In loosely packed media, we find that the stress relaxation curve can only be fitted using two distinct relaxation time scales, which indicates that this process is governed by two physical mechanisms. Comparison of the relaxation time scales to literature values indicates that stress relaxation is most likely governed by i) mechanical rearrangement of particles and ii) fluid redistribution inside the medium. In samples with higher packing fraction, we only find one relaxation time scale. In this case, the connectivity of the contact network inside the granular medium is larger and particle movement is suppressed.

The results of our study show that the notion of ultimate tensile strength, as measured by continuous extension experiments, does not provide a complete description of a granular material. This is especially true when the deformation of the medium is stopped or when deformation time scales are comparable to the relaxation time scales found in this study. In such cases, the actual strength of the medium might be lower than suggested by experiments, because the stress state of the medium is altered by the phenomena of fluid redistribution and particle rearrangement. This has e.g. implications for the assessment of slope stability, where the critical slope could be lower when tensile stresses are reduced.

\section*{Acknowledgements}
\label{sec:Conclusions}

We acknowledge financial support from the European Research Council (ERC) Advanced Grant n. 319968 FlowCCS. The technical assistance of Daniel Breitenstein in constructing the experimental apparatus is greatly appreciated. The final publication is available at Springer via https://doi.org/10.1007/s10035-016-0673-6.

\section*{Conflict of interest}
\label{sec:Conflict of interest}

The authors declare that they have no conflict of interest.

%

\bibliographystyle{spmpsci}      
\bibliography{Manuscript}   

\begin{thebibliography}{10}
\providecommand{\url}[1]{{#1}}
\providecommand{\urlprefix}{URL }
\expandafter\ifx\csname urlstyle\endcsname\relax
  \providecommand{\doi}[1]{DOI~\discretionary{}{}{}#1}\else
  \providecommand{\doi}{DOI~\discretionary{}{}{}\begingroup
  \urlstyle{rm}\Url}\fi

\bibitem{Carr1967}
Carr, J.F.: Tensile strength of granular materials.
\newblock Nature \textbf{213}(5081), 1158--1159 (1967)

\bibitem{Fisher1926}
Fisher, R.A.: On the capillary forces in an ideal soil; correction of formulae
  given by w. b. haines.
\newblock The Journal of Agricultural Science \textbf{16}, 492--505 (1926)

\bibitem{Francois2012}
Fran\c{c}ois, D., Pineau, A., Zaoui, A.: Mechanical Behaviour of Materials.
\newblock Springer (2012)

\bibitem{German2014}
German, R.M.: Coordination number changes during powder densification.
\newblock Powder Technology \textbf{253}(0), 368 -- 376 (2014)

\bibitem{Haines1925}
Haines, W.B.: Studies in the physical properties of soils: Ii. a note on the
  cohesion developed by capillary forces in an ideal soil.
\newblock The Journal of Agricultural Science \textbf{15}, 529--535 (1925)

\bibitem{Hartley2003}
Hartley, R.R., Behringer, R.P.: {Logarithmic rate dependence of force networks
  in sheared granular materials.}
\newblock Nature \textbf{421}(6926), 928--931 (2003)

\bibitem{Herminghaus2005}
Herminghaus, S.: Dynamics of wet granular matter.
\newblock Advances in Physics \textbf{54}(3), 221--261 (2005)

\bibitem{Hornbaker1997}
Hornbaker, D.J., Albert, R., Albert, I., Barab\'{a}si, A.L., Schiffer, P.:
  {What keeps sandcastles standing ?}
\newblock Nature \textbf{387}, 765 (1997)

\bibitem{Iveson2001}
Iveson, S.M., Litster, J.D., Hapgood, K., Ennis, B.J.: Nucleation, growth and
  breakage phenomena in agitated wet granulation processes: a review.
\newblock Powder Technology \textbf{117}(1–2), 3 -- 39 (2001).
\newblock Granulation and Coating of Fine Powders

\bibitem{Kim2003}
Kim, T.H., Hwang, C.: Modeling of tensile strength on moist granular earth
  material at low water content.
\newblock Engineering Geology \textbf{69}(3–4), 233 -- 244 (2003)

\bibitem{Kohonen2004}
Kohonen, M.M., Geromichalos, D., Scheel, M., Schier, C., Herminghaus, S.: On
  capillary bridges in wet granular materials.
\newblock Physica A: Statistical Mechanics and its Applications
  \textbf{339}(1–2), 7 -- 15 (2004).
\newblock Proceedings of the International Conference New Materials and
  Complexity

\bibitem{Kohonen1999}
Kohonen, M.M., Maeda, N., Christenson, H.K.: Kinetics of capillary condensation
  in a nanoscale pore.
\newblock Physical Review Letters \textbf{82}, 4667--4670 (1999)

\bibitem{Kristensen1985}
Kristensen, H., Holm, P., Schaefer, T.: Mechanical properties of moist
  agglomerates in relation to granulation mechanisms part ii. effects of
  particle size distribution.
\newblock Powder Technology \textbf{44}(3), 239 -- 247 (1985)

\bibitem{Labajos2006}
Labajos-Broncano, L., Antequera-Barroso, J., Gonz\'{a}lez-Mart\'{i}n, M.,
  Bruque, J.: An experimental study about the imbibition of aqueous solutions
  of low concentration of a non-adsorbable surfactant in a hydrophilic porous
  medium.
\newblock Journal of Colloid and Interface Science \textbf{301}(1), 323 -- 328
  (2006)

\bibitem{Lambert2008}
Lambert, P., Chau, A., , Delchambre, A., R\'{e}gnier, S.: Comparison between
  two capillary forces models.
\newblock Langmuir \textbf{24}(7), 3157--3163 (2008)

\bibitem{Lian2016}
Lian, G., Seville, J.: The capillary bridge between two spheres: New
  closed-form equations in a two century old problem.
\newblock Advances in Colloid and Interface Science \textbf{227}, 53 -- 62
  (2016)

\bibitem{Lian1993}
Lian, G., Thornton, C., Adams, M.J.: A theoretical study of the liquid bridge
  forces between two rigid spherical bodies.
\newblock Journal of Colloid and Interface Science \textbf{161}(1), 138 -- 147
  (1993)

\bibitem{Lu2007}
Lu, N., Wu, B., Tan, C.: Tensile strength characteristics of unsaturated sands.
\newblock Journal of Geotechnical and Geoenvironmental Engineering
  \textbf{133}(2), 144--154 (2007)

\bibitem{Mani2013}
Mani, R., Kadau, D., Herrmann, H.: Liquid migration in sheared unsaturated
  granular media.
\newblock Granular Matter \textbf{15}(4), 447--454 (2013)

\bibitem{Mani2012}
Mani, R., Kadau, D., Or, D., Herrmann, H.J.: Fluid depletion in shear bands.
\newblock Phys. Rev. Lett. \textbf{109}, 248,001 (2012)

\bibitem{Mani2015}
Mani, R., Semprebon, C., Kadau, D., Herrmann, H.J., Brinkmann, M., Herminghaus,
  S.: Role of contact-angle hysteresis for fluid transport in wet granular
  matter.
\newblock Phys. Rev. E \textbf{91}, 042,204 (2015)

\bibitem{Mitarai2006}
Mitarai, N., Nori, F.: Wet granular materials.
\newblock Advances in Physics \textbf{55}(1-2), 1--45 (2006)

\bibitem{Pierrat1998}
Pierrat, P., Agrawal, D.K., Caram, H.S.: Effect of moisture on the yield locus
  of granular materials: theory of shift.
\newblock Powder Technology \textbf{99}(3), 220 -- 227 (1998)

\bibitem{Pierrat1997}
Pierrat, P., Caram, H.S.: Tensile strength of wet granula materials.
\newblock Powder Technology \textbf{91}(2), 83 -- 93 (1997)

\bibitem{Rumpf1962}
Rumpf, H.: Agglomeration, pp. 379--418.
\newblock New York, Interscience (1962)

\bibitem{Scheel2009}
Scheel, M.: {Experimental investigations of the mechanical properties of wet
  granular matter}.
\newblock Ph.D. thesis, Georg-August-Universit\"{a}t G\"{o}ttingen (2009)

\bibitem{Scheel2008NM}
Scheel, M., Seemann, R., Brinkmann, M., {Di Michiel}, M., Sheppard, a.,
  Breidenbach, B., Herminghaus, S.: {Morphological clues to wet granular pile
  stability.}
\newblock Nature Materials \textbf{7}(3), 189--93 (2008)

\bibitem{Scheel2008JP}
Scheel, M., Seemann, R., Brinkmann, M., {Di Michiel}, M., Sheppard, A.,
  Herminghaus, S.: {Liquid distribution and cohesion in wet granular assemblies
  beyond the capillary bridge regime}.
\newblock Journal of Physics: Condensed Matter \textbf{20}(49), 494,236 (2008)

\bibitem{Schiffer2005}
Schiffer, P.: {A bridge to sandpile stability}.
\newblock Nature Physics \textbf{1}, 21--22 (2005)

\bibitem{Schubert1975}
Schubert, H., Herrmann, W., Rumpf, H.: Deformation behaviour of agglomerates
  under tensile stress.
\newblock Powder Technology \textbf{11}(2), 121 -- 131 (1975)

\bibitem{Seeman2001}
Seemann, R., M\"{o}nch, W., Herminghaus, S.: Liquid flow in wetting layers on
  rough substrates.
\newblock Europhysics Letters, \textbf{55}, 698--704 (2001)

\bibitem{Takenaka1981}
Takenaka, H., Kawashima, Y., Hishida, J.: The effects of interfacial physical
  properties on the cohesive forces of moist powder in air and in liquid.
\newblock Chemical and Pharmaceutical Bulletin \textbf{29}(9), 2653--2660
  (1981)

\bibitem{Turba1964}
Turba, E., Rumpf, H.: Zugfestigkeit von pre\ss lingen mit vorwiegender bindung
  durch van der waals-kr\"{a}fte und ihre beeinflussung durch
  adsorptionsschichten.
\newblock Chemie Ingenieur Technik \textbf{36}(3), 230--240 (1964)

\bibitem{Utter2004}
Utter, B., Behringer, R.: Transients in sheared granular matter.
\newblock The European Physical Journal E \textbf{14}(4), 373--380 (2004)

\bibitem{Willett2000}
Willett, C.D., Adams, M.J., Johnson, S.A., Seville, J.P.K.: {Capillary Bridges
  between Two Spherical Bodies}.
\newblock Langmuir \textbf{16}(24), 9396--9405 (2000)

\end{thebibliography}

%
%

\end{document}